# Short-period stellar activity cycles with *Kepler* photometry

Oleksiy V. Arkhypov, Maxim L. Khodachenko[1] and Helmut Lammer Space Research Institute, Austrian Academy of Sciences, Schmiedlstrasse 6, A-8042 Graz, Austria

oleksiy.arkhypov@oeaw.ac.at

Manuel Güdel, Theresa Lüftinger,

and

Colin P. Johnstone

Institute of Astronomy, University of Vienna, Türkenschanzstrasse 17, 1180 Vienna, Austria

---

[1] Skobeltsyn Institute of Nuclear Physics, Moscow State University, Moscow, Russia

# ABSTRACT

We study the short-periodic component of stellar activity with a cycle periods $P_{cyc} < 10^3$ days using the *Kepler* mission photometry of fast-rotating ($1 < P < 4$ days) stars with spectra of M4V to F3V. Applying the originally developed two non-spectral methods, we measured the effective period of stellar cycles in 462 objects. The obtained results are in accordance with previous measurements by Vida et al. (2014), do not seem to result from a beating effect. The performed measurements of $P_{cyc}$ cluster in a specific branch which covers the previously unstudied region in the Saar-Brandenburg (1999) diagram, and connects the branch of inactive stars with the area populated by super-active objects. It is shown that the formation of the discovered branch is due to the cu-quenching effect, which saturates the magnetic dynamo and decreases the cycle periods with the increase of inverted Rossby number. This finding is important in the context of the discussion on catastrophic quenching and other heuristic approximations of the non-linear α-effect.

*Subject headings:* stars: activity — stars: magnetic field — (stars:) starspots — stars: statistics

## 1. Introduction

Stellar activity is a key factor of space weather and a valuable information channel on processes in stellar interiors (e.g., differential rotation, circulation and convection). At the same time, the connection of these processes with surface activity is poorly known. Various theories of magnetic dynamos need further observational tests and verification. That is why the empirical approach plays an important role in studies of stellar activity phenomenology. The matter of this approach consists in the search for the relation between physical parameters of stars and their activity parameters (e.g., the cycle periods $P_{cyc}$ and the average activity levels $A$) and comparison of that with theoretical predictions.

An interesting view on stellar activity is given in Fig. 1 (reproduced from Saar & Brandenburg, 1999, Fig. 5; Herein after SB diagram). It shows that the stars in relation to their $\omega_{cyc}/\Omega$ and the inversed Rossby number $Ro^{-1} = 2\tau_{MLT}\Omega$ are clustered in 3 branches: inactive (I), active (A) and super-active (S) objects. Here $\omega_{cyc} = 2\pi/P_{cyc}$ is the angular velocity of the magnetic cycle, $\Omega = 2\pi/P$ is the angular velocity of the stellar rotation, $P$ is the rotation period of star and $\tau_{MLT}$ is the turnover time according to the mixing length theory (MLT; Bohm-Vitense, 1958).

Unfortunately, the limited sample of the considered stars in the SB diagram does not cover the region of $\log(\omega_{cyc}/\Omega) > -2$ at $Ro^{-1} > 1$, which still has to be explored. The uncovered region corresponds to the fast rotating stars and the shortest stellar activity cycles. In the present paper, we investigate this unstudied area using new data from the *Kepler* mission.

Our study is focused on the stars with $1 < P < 4$ days and shortest scale of activity cycles $P_{cyc} < 10^3$ days. A few similar objects with short activity cycles were initially investigated using Earth-based data (Vida & Olah, 2006; Olah et al., 2009; Vida et al., 2013). The previous analysis of *Kepler* light-curves also confirmed the existence of stellar cycles with $300 < P_{cyc} < 600$ days (Vida & Olah, 2014; Vida et al, 2014). Unfortunately, the laborious methods of light-curve analysis applied in Vida & Olah (2014) and Vida et al. (2014) have limited the number of finds to only 9 objects. In this work, we propose automatized analysis methods which we apply to hundreds of stars.

Mathur et al. (2014) argue that at least in two cases, the short-cycle-type *($P_{cyc}$ =89 and

463 days) phenomena could be interpreted as a beating of two frequencies from two spots rotating at different latitudes under the condition of a significant differential rotation of stars. However, such beating-effect should have a large dispersion of its periods connected with different latitudinal separation of spots on different stars (in spite of their general similarity, e.g., stellar type). On the contrary, the activity cycles of similar stars would show clustering around certain $P_{cyc}$, in accordance with the true cycle periods, measured using the butterfly-method, which is insensitive to the beating effect (Vida & Szabo, 2014). In the present paper, we test this expectation. Moreover, we compare our estimates of $P_{cyc}$ with theoretical predictions for an αω-dynamo.

Since the short cycles of activity are associated with the I-branch in SB diagram (Fig. 1), which is supposedly related to the tachocline dynamo (Bohm-Vitense, 2007), our study could also be useful for developing a better understanding of the major (i.e., tachocline- related) solar activity.

## 2. Stellar set

The high-precision light-curves of 513 Main Sequence stars from the *Kepler* mission (NASA Exoplanet Archive, 2013) of M1V to F5V spectral classes and clear spot variability were selected for our analysis (Table 1) using the reference catalog of 12 000 main-sequence *Kepler* stars (Nielsen et al., 2013).

Only the stars with rotation periods between f and 4 days were considered. The lower limit is due to the restriction of the used reference catalog (Nielsen et al., 20f 3). The upper limit was chosen to get the stars with the maximal number of rotations during the observation time, while still providing a large enough sample of stars. Note that most of the defined stellar period values in the used catalogue my be assumed to be correct since they are in good agreement with those given in other sources. In particular, the comparison of the data in the reference catalog (Nielsen et al., 2013) and the catalog of McQuillan et al. (2014) shows that differences of more than 10% in the rotation period values takes place only in 2.93% of cases.

In practice, we cannot find stars in the reference star catalog (Nielsen et al., 2013) cooler than 3504 K which have the rotation periods within the above specified interval of values. On the other hand, to avoid the contamination of the stellar sample by pulsating variables, we limit our study to stars with the effective temperature $T_{eff} < 6750$ K. In particular, we found that about 80% of stars with $T_{eff} = 7000 \pm 250$ K in the reference catalog (Nielsen et al., 2013) show signs of pulsations (quasi-stable amplitudes, multiple peaks in periodograms, oscillations with too short periods of ~ 0.1 day, etc.).

To exclude binary systems and pulsating stars with typically stabile or periodically varying amplitude (not related with spot activity), we select stars that show irregular and gradual variations in the amplitude of the light-curve oscillations (e.g. in Fig. 2a).

As an additional selection criterion, we also used the domination of rotational harmonics in the quarterly Lomb-Scargle periodograms of a star similar to the case shown in Fig. 2c. To exclude the interference caused by stellar companions and pulsations, we discarded the periodograms that contain noticeable extra-peaks (not related with the rotational harmonics *P, P/2* and P/3) and noise at periods shorter than *2P*. We also tried to avoid stars with too noisy light-curves, i.e., when the dispersion of the light-curve in the narrow time intervals (<< *P)* is larger than 10% of the amplitude of the first rotational harmonic.

For each star, we processed data from quarters 0 to 16 in the *Kepler* archive (http://exoplanetarchive.ipac.caltech.edu/applications/ETSS/KeplerJndex.html) with long cadences. Each series of quarterly data contains 60,000 measurements of the radiation flux *F* (PDCSAP_FLUX) per star corrected for instrumental and environmental effects.

### 3. Methods

Searching for the long-term variations in *Kepler* light-curves (i.e., in the radiation flux *F* (PDCSAP_FLUX) in the aperture as a function of time *t)* is not a trivial task because of significant instrumental trends and flux jumps from one quarter to another. So far two methods to deal with this problem have been applied: 1) the analysis of the temporal variation of the standard deviation or variance of the light-curve for subseries (Mathur et al. 2013); and 2) the search for changes in the rotation period due to the differential rotation and the butterfly-effect or the periodical drift of the stellar spots in latitude during the activity cycles (Vida et al., 2014). Both methods require individual tuning of data and spectral analysis in every particular case. That is why they were used for the studies of only a small number (a few dozens) of stars. In our approach based on the same *Kepler* data, to increase the stellar statistics, we analyzed every rotation of the star using a new activity index, introduced below. Since the available data from the *Kepler* mission cover 4 years, our study is focused on the short-period components of stellar activity. To analyze these features, we develop two processing methods which are specialized to high-frequency oscillations of stellar activity.

### 3.1. Activity index

The classical indices of solar activity for broadband emission, such as the total sunspot number and sunspot area, are inapplicable in the case of unresolved stellar disks. In that respect, the activity cycles have usually been studied as a variability of smoothed stellar magnitudes or narrow-band emission fluxes, without any rotational modulation (e. g., Olah et al., 2009 and therein). Unfortunately, unavoidable instrumental artifacts (trends and jumps like in Fig. 2) in the *Kepler* photometrical series makes it impossible to study some long-time stellar variability.

At the same time, information on activity cycles in the *Kepler* photometry can be deduced from analysis of just the rotational variability of the stellar light-curves. The direct search for oscillations in spot-related rotation periods (Vida et al.2014) is a heavy way to survey numerous stars. An easier approach consists of the analysis of the amplitude of rotation modulation. For example, Mathur et al. (2013) studied the temporal variation of the standard deviation of light-curves in a sliding time-window. This method is however sensitive to photon noise, sporadic flares, pulsations, and micro-variability.

It is natural to expect that the most reliable manifestation of starspots is the amplitude of the first harmonic $A_1$ of rotational modulation in a light-curve. This parameter is practically insensitive to the photon noise and it minimally depends on short-lived flares and short-period pulsations in contrast with other rotational harmonics. Usually, the value $A_1$ can be measured with maximum accuracy because the light integration over the stellar disk reduces the amplitude of higher harmonics progressively with the increasing harmonic number.

Since the first harmonic is maximal, the value $A_1^2$ is a main contributor to variance $<(F - (F))^2>$ of the stellar light-curve, where $<...>$ means the averaging in time. On the other hand, this variance is the sum of the sub-variances $\langle(\delta F_i)^2\rangle$ from particular star spots with $i$-numbers:

$$\langle(F - \langle F\rangle)^2\rangle = \sum_{i=1}^{N_s}\langle(\delta F_i)^2\rangle = N_s \xi, \qquad (1)$$

where $\xi$ is the average value of sub-variance from individual starspot and $N_s$ is the total number of spots on the star. Hence, it can be supposed that $A_1^2 \propto \langle(F - \langle F\rangle)^2\rangle \propto N_s$.

We check the expectation $A_1^2 \propto N_s$ by comparing the composite of measurements of total solar irradiance (TSI) 1978-2003 (ftp://ftp.ngdc.noaa.gov/STP/SOLAR_DATA/SOLARJRRADIANCE/) with the daily total sunspot number (*W;* http://sidc.oma.be/silso/infossntotdaily). Using the TSI measurements, we calculate the time sequence of values of $A_1^2/A_0^2$, where $A_0^2$ is the squared zero harmonic, i.e., the squared average value $<F>^2$. Figure 3 shows the sequences of values of $A_1^2/A_0^2$ and *W*. The sunspot number *W* is averaged over the same time intervals as those used for the calculation of $A_1^2/A_0^2$ (i.e., over 27.2753 days of the Carrington solar rotation period). One can see in Figure 3, three cycles of activity in both indices. However, the index $A_1^2/A_0^2$ shows more clearly the high-frequency fluctuations, which are the subject of our study.

Figure 4 demonstrates the relation between both indices $A_1^2/A_0^2$ and *W*. Since the value of $A_1^2/A_0^2$ describes an asymmetry in the longitudinal distribution of sunspots rather than the total spot number, the correlation is not high but significant (0.63±0.03). The observed scatter of the index $A_1^2/A_0^2$ in Figure 4 is mainly a result of varying distribution of the activity regions in longitude. The averaging of $A_1^2/A_0^2$ within narrow intervals of *W* (black squares in Figure 4) reveals a quasi-linear increase with the sunspot number. The linear regression $A_1^2/A_0^2 = GW + Q$ (dashed line in Fig. 4) has the coefficient $G = 1.17 \pm 0.08$ about unity. This confirms the prediction of the statistical relation $A_1^2 \propto N_s$. That is why we employ the index $A_1^2$ for our study of stellar activity. Henceforth the normalization coefficient $A_0^{-2}$ is omitted, because we deal with the relative flux $F/<F>$, hence $A_o = 1$.

### 3.2. Light-curve processing

The stellar light-curves were prepared in a special way (described below) to reduce the non-rotational stellar variability. To calculate the activity index $A_1^2$ with maximal time resolution, we divided the light-curves $F(t)$ into consecutive fragments which have durations of the stellar rotation period each (Fig. 5a).

To remove stellar flares in a one-period light-curve (Fig. 5a), we calculate an average flux value, $<F>$, and the dispersion, $\sigma$, at the same rotational phases, i.e. for $F(t - P)$, $F(t)$ and $F(t + P)$. If the condition $[F(t) - F(t - P)]/\sigma > 1$ or $[F(t) - F(t + P)]/\sigma > 1$ is satisfied, then we replace the measured $F(t)$ with the value $[F(t - P) + F(t + P)]/2$. Figure 5b shows the example of such flare removal. After this we remove the linear trend, so that $F(t) = F(t + P)$. To provide equidistant counts of $F$, we filled the small ($< 0.1P$) gaps in the light-curves using linear interpolation. The periods with larger gaps were ignored. Then each of the one-period light-curve segments was normalized to its mean flux.

The next step was the calculation of the activity index. For each one-period light-curve, we calculate an individual squared amplitude $A_1^2$ of the first rotational spectral harmonic in the discrete Fourier transform:

$$A_1^2 = a_1^2 + b_1^2, \qquad (2)$$

$$a_1 = \frac{1}{N} \sum_{i=1}^{N} F_i \cos \frac{2\pi i}{N}, \qquad (3)$$

$$b_1 = \frac{1}{N} \sum_{i=1}^{N} F_i \sin \frac{2\pi i}{N}, \qquad (4)$$

where $N$ is the total number of flux readings ($F_i$) in the one-period light-curve and $i$ is the flux reading number in the one-period light-curve.

We do not measure stellar rotation periods, but use the published average values (Nielsen et al., 2013). Typically, differential rotation and the variability of spot latitudes could cause a few percent variation $\Delta P$ in $P$ (Reinhold et al., 2013). The analyzed set finally consists of 513 objects. Figure 6 shows that the corresponding errors $1 - \delta$ in the activity index $A_1^2$, where $\delta = <A_1^2(P + \Delta P)/A_1^2(P)>$, are negligible ($|1 - \delta| < 10\%$) at $\Delta P/P < 0.1$. Note that in the course of visual control of the light-curves of considered objects two stars (KIC11349386 and KIC12066633) were excluded from the analyzed set because of an erroneously estimated period ($P/2$ instead of $P$) in the catalog. A comparison of the used periods $P$ in our stellar set with another catalog (McQuillan et al., 2014) revealed no errors $|\Delta P|/P > 0.072$.

With the above described approach, we obtained chronological sequences of values of $A_1^2$. Figure 7 shows that there are cyclic oscillations of the stellar activity index. Note that the time here is represented in differential barycentric Julian days (ΔJD), counted from the observation starting time. The numerous spectral peaks and the general increasing trend of spectral power toward longer periods of harmonics (right plots in Fig. 7) make the estimation of the cycle period $P_{cyc}$ a non-trivial task. Any type of spectral smoothing applied here will increase the estimates of the detectable minimal periods and therefore distort the real cycle statistics. Usually the activity cycles with longest periods dominate in the spectrum, whereas the subject of our study is the relatively short-period cycles. Moreover, for long periods with low harmonic number $m<10$, the spectral resolution $\Delta P_1 = P_{max}/m$, is the same order of magnitude as the maximal detectable period $P_{max}$, which is the duration of light-curve. This is why the traditional Fourier analysis methods are inadequate for the considered problem, and below we apply an alternative - non-spectral-methods in our study of the short-period cycles.

### 3.3. Autocorrelation method (ACM)

The autocorrelation method (ACM) considers the analyzed function, e.g., the fluctuation of the variance $A_1^2$, $\Delta I(t) = A_1^2(t) - <A_1^2>$, as a stochastically modulated oscillation

$$\Delta I = K(t) \sin(\omega_{cyc} t + \varphi), \quad (5)$$

where $K(t)$ is a time-dependent factor of the stochastic modulation caused by the emergence of active regions on the stellar surface, $\omega_{cyc} = 2\pi/P_{cyc}$ is a typical cycle frequency for the stellar variability (i.e., activity cycle), $t$ is a time, and $\varphi$ is a phase. Considering $K$ as a fluctuation with an average value of $<K> = 0$, one can find an autocorrelation of $\Delta I$ at the lag $\Delta t$:

$$r(\Delta t) = \frac{\langle \Delta I(t) \Delta I(t + \Delta t) \rangle}{\langle \Delta I \rangle^2}. \quad (6)$$

Substituting Eq. (5) in (6), one can obtain

$$r(\Delta t) = \frac{\langle K(t) \sin(\omega_{cyc} t + \phi) K(t + \Delta t) \sin(\omega_{cyc}(t + \Delta t)\phi) \rangle}{\langle \Delta A^2 \rangle^2}. \quad (7)$$

Using the trigonometric formula $\sin\alpha \sin\beta = [\cos(\alpha-\beta) - \cos(\alpha+\beta)]/2$, one can replace $\sin(\omega_{cyc} t + \varphi)$ $\sin[\omega_{cyc}(t + \Delta t) + \varphi]$ with $[\cos(-\omega_{cyc}\Delta t) - \cos(2\omega_{cyc} t + \omega_{cyc} \Delta t + 2\varphi)]/2$ in Eq. (7), where the second oscillating term $\cos(2\omega_{cyc} t + \omega \Delta t + 2\varphi)$ gives a negligible (less than percent of the first term $\cos(-\omega \Delta t)$) contribution in autocorrelation after averaging. Moreover, it follows from Eq. (5) that

$$\langle \Delta I^2 \rangle = \langle K^2 \rangle \langle \sin^2(\omega_{cyc} t + \phi) \rangle = \langle K^2 \rangle/2 \tag{8}$$

for the non-correlated values $K^2$ and $\sin^2(\omega_{cyc} t + \varphi)$. Hence, it follows from Eqs. (6)-(8) that

$$r(\Delta t) = \eta \cos(\omega_{cyc} \Delta t), \tag{9}$$

where $\eta \equiv \langle K^2 \rangle$ is the autocorrelation of $K$.

Based on the exponential approximation used in Welsch et al. (2012) for the autocorrelation coefficient of the temporal evolution of solar activity patterns, we suppose for short non-zero lags that $r = (1 - c) \exp(-\Delta t/\tau) \cos(\omega_{cyc} \Delta t)$, where $\tau$ is the time scale of $\eta$, and $c$ is a small constant which corresponds to the decorrelation jump between $\Delta t/P = 0$ and 1 (due to photon noise, micro-flares and granulation variability with timescales shorter than $P$). Using a Taylor series approximation for $c \ll 1$ and short lags $\Delta t \ll 2\pi/\omega_{cyc}$, (i.e. $\ln(1 - c) \approx -c$ and $\ln[\cos(\omega_{cyc} \Delta t)] \approx -0.5(\omega_{cyc} \Delta t)^2$), we obtain

$$\ln(r) \approx -c - \frac{\Delta t}{\tau} - \frac{1}{2}\omega_{cyc}^2 (\Delta t)^2. \tag{10}$$

On the other hand, we approximate the logarithm of $r(\Delta t)$ using a quadratic polynomial at shortest lags

$$\ln(r) \approx c + b\,\Delta t + a\,(\Delta t)^2. \tag{11}$$

Figure 8 shows that this approximation works well, sometimes up to $\Delta t/P \sim 10$. Here $r \equiv r_1$ is the autocorrelation coefficient using the first harmonic in the chronological sequence of $A_1^2$ for the standard set of time lags $\Delta t = kP$, where $k = 1, 2, 3, 4$:

$$r_1(\Delta t) = \frac{1}{N} \frac{\sum_{i=0}^{N-1}\{[A_1^2(t) - \langle A_1^2 \rangle][A_1^2(t + \Delta t) - \langle A_1^2 \rangle]\}}{\sigma_1 \sigma_2}, \tag{12}$$

where

$$\sigma_1 = \sqrt{\frac{1}{N} \sum_{j=0}^{N-1} [A_1^2(t) - \langle A_1^2(t) \rangle]^2}, \tag{13}$$

$$\sigma_2 = \sqrt{\frac{1}{N} \sum_{k=0}^{N-1} [A_1^2(t + \Delta t) - \langle A_1^2(t + \Delta t) \rangle]^2}. \tag{14}$$

Here $\langle A_1^2(t) \rangle = (1/N)\sum_{i=0}^{N-1} A_1^2(t)$ and $\langle A_1^2(t + \Delta t) \rangle = (1/N)\sum_{i=0}^{N-1} A_1^2(t + \Delta t)$ are the average values and $N$ is the number of pairs $A_1^2(t)$ and $A_1^2(t + \Delta t)$, which were found in the chronological sequence.

For calculating the coefficients *a, b, c* and their standard errors, we use the least square method and the minimal set of the lags $\Delta t/P = 1, 2, 3$ and $4$ to minimize the effect of ignoring the terms with $\Delta t^3$, $\Delta t^4$, etc., in Equation (11). Using the known Gauss method, one can calculate $a=d_a/d$, $b=d_b/d$ and $c=d_c/d$, where $d_a=v_{14}(v_{22}v_{33}-v_{23}v_{32})-v_{12}(v_{24}v_{33}-v_{34}v_{23})+v_{13}(v_{24}v_{32}-v_{34}v_{22})$; $d_b=v_{11}(v_{24}v_{33}-v_{23}v_{34})-v_{13}(v_{21}v_{34}-v_{31}v_{24})$; $d_c=v_{11}(v_{22}v_{34}-v_{24}v_{32})-v_{12}(v_{21}v_{34}-v_{31}v_{24})+v_{14}(v_{22}v_{33}-v_{23}v_{32})$; $d=v_{11}(v_{22}v_{33}-v_{23}v_{32})-v_{12}(v_{21}v_{33}-v_{31}v_{23})+v_{13}(v_{21}v_{32}-v_{31}v_{22})$. Here the values $v_{j,k}$ are the result of summation of a set of measurements of $r_1(\Delta t)$ at different autocorrelation lags $\Delta t = P$, $2P$, $3P$, $4P$: $v_{11} = \Sigma \Delta t^4$; $v_{12} = v_{21} = \Sigma \Delta t^3$; $v_{13} = v_{22} = v_{31} = \Sigma \Delta t^2$; $V_{14} = \Sigma \ln[r(\Delta t)]\Delta t^2$; $v_{23} = v_{32} = \Sigma \Delta t$; $v_{24} = \Sigma \Delta t \ln[r(\Delta t)]$; $v_{33} = 4$ is the lag number; $v_{34} = \Sigma \ln[r(\Delta t)]$. The standard errors of *a, b* and *c* are respectively: $\sigma_a = \varepsilon/P_a^{1/2}$; $\sigma_b = \varepsilon/P_b^{1/2}$; $\sigma_{cc} = \varepsilon/P_c^{1/2}$, where $\varepsilon^2=\delta/(v_{33}-3)$, $\delta=\Sigma\{\ln[r(\Delta t)]-a\Delta t^2-b\Delta t-c\}^2$; $P_a=d/(v_{22}v_{33}-v_{32}v_{23})$; $P_b=d/(v_{11}v_{33}-v_{31}v_{13})$; $P_c=d/(v_{11}v_{22}-v_{21}v_{12})$.

Comparing Equations (10) and (11), we estimate the coefficient $a=-0.5\omega_{cyc}^2$, and therefore, the angular frequency $\omega_{cyc}=(-2a)^{1/2}$ of the stellar cycle. The corresponding period $P_{ACM}= 2\pi/\omega_{cyc}$ describes the short-period component of stellar variability, because mainly the fast oscillations of $\Delta I$ control the autocorrelation *r* at shortest lags $\Delta t$. The error of $P_{ACM}$ was calculated as $\sigma_{ACM}=0.5|2\pi(\omega_1^{-1}-\omega_2^{-1})|$, where $\omega_{1,2}=[-2(a\pm\sigma_a)]^{1/2}$.

We test the autocorrelation method with the following model. Since the used index of activity $A\backslash$ is based on the first harmonic of the one-period light-curve $F(t)$, we can consider only this harmonic. As a rule, this is the dominant component of the whole light-curve. We simulate it as a superposition of sinusoidal signals caused by individual starspots and faculae. Such a general approach enables the reduction of the number of model parameters (i.e. inclination angle of stellar rotation axis, latitude, effective area and contrast of a spot), which are implicitly included in the amplitude of the sinusoid. In this way,

$$F(t_i) = 1 + \sum_{j=1}^{N_s} G_j E_j(t_i) \sin[\Omega t_i + \varphi_j], \qquad (15)$$

where $t_i$ is the discrete time count, $N_s$ is the total number of starspots, which control the whole light-curve; $j$ is a number of particular starspot; $G_j = A_S$ RND is an amplitude of the j-th spot contribution to the light-curve ($A_s$ is the maximal spot area relative the total stellar disk, 0 < RND < 1 is a random value), and $E_j(t_i)=\exp[—(U — S_j)/\tau]$ at $t_i > S_j$ and $E_j(t_i)= 0$ at $t_i < S_j$ is the coefficient of spot evolution with timescale. Here, $S_j$ stands for the spot appearance time, $\Omega$ is the angular frequency of stellar rotation, and $\varphi_j = 2\pi$ RND is the random phase describing the spot's location in longitude. The activity cycle modulates the frequency of the spot appearance in the form of a sinusoid with period $P_{cyc}$ as

$$S_j = P_{cyc}\text{Int}\left(\frac{T_{tot}\text{RND}}{P_{cyc}}\right) + \Delta S, \qquad (16)$$

where Int denotes the integer part of the value in brackets, $T_{tot}$ is the total duration of the light-curve, and $\Delta S = \mu P_{cyc}/\pi$ with an equiprobable $\mu=\arccos[2(\text{RND} — 0.5)]$ or $\mu=2\pi— \arccos[2(\text{RND} - 0.5)]$.

We take realistic values of the model parameters based on present day knowledge. The typical area of non-polar (as the most effective modulators) spots on fast-rotating (1.2 < P < 2.9 days) main-sequence stars is estimated photometrically as $\sim 10^{-2}$ in units of the stellar disk area (Frasca et al. 2011; Frohlich et al. 2012). Although the faculae were not included in the models used for these estimations, they have nevertheless an influence on the area of the modeled spots through the light-curve fitting. Therefore, the estimated effective spot area naturally includes the effect of the faculae.

However, the solar spots typically are much smaller $\sim 10^{-4}$ (Allen, 1973). It is very probable that the giant stellar "spots" are spot clusters (regions or complexes of activity) with unresolved structure.  That is why we use a compromise estimate $A_s=10^{-3}$. The ratio of new spot groups number during a year to the average total sunspot number in the solar disk is about $\zeta = 5.0$ when $W \sim 100$ at the typical maximum (Allen, 1973). Hence, the number of new spots during the time of the *Kepler* observation $T_{tot} = 1500$ days is $N_s > \zeta W T_{tot}/(365.24 \text{ days}) \approx 2 \times 10^3$. Thus, we take $N_s = 3 \times 10^3$. By comparison of Equations (10) and (11), we estimate that $\tau = -1/b \approx 200$ days and $\tau \sim 10$ days for the "cold" ($T_{eff} \sim 3500$ K) and "hot" ($T_{eff} \approx 6700$ K) objects in our stellar set, respectively (Arkhypov et al., 2015). The cycle period $P_{cyc}$ is suggested to be 50 or 300 days in accordance with preliminary estimates for hot and cold stars, respectively. The period of stellar rotation $P = 3$ days is taken as a typical value in our stellar set.

Using this model, we synthesized 576 light-curves for testing of the above specified methods for $P_{cyc}$ measurement. An example synthetic light-curves (Figure 9a,i) reproduces the oscillations with a variable amplitude, which is typical for real light-curves (Figure 2). The behavior of the corresponding activity index (Figure 9c,k) looks similar to the real cases in Figures 7e and 10i. There is a resemblance between synthetic power spectra of $A_1^2$ in Figures 9f,n and real spectra in 9f,l. The autocorrelation curves (Figure 9d,l) exhibit a behavior similar to the real one (Figure 10d,j). Note that the histograms of the estimated cycle period $P_{ACM}$ in the simulations are peaked near the model value $P_{cyc}$ (pointed line in Figure 9g,o). In fact, the average values $\langle \log(P_{ACM}) \rangle$ correspond to $P_{ACM} = 44.0 \pm 1.7$ and $304.4 \pm 14.3$ days, which are close to the model input parameter $P_{cyc} = 50$ and 300 days. Therefore, despite considerable dispersion of individual estimates of $P_{ACM}$, the average value $\langle \log(P_{ACM}) \rangle$ may be used as the correct one.

The application of the autocorrelation method to real stars (Figures 7 and 10) confirms that the value $P_{ACM}$ acceptably approximates the Fourier spectral peaks of short-period oscillations even on the background of dominating long cycles in power spectra of $A_1^2(t)$ (Figures 7d and 10 1).

Moreover, we tried to apply the autocorrelation method to the composite of measurements of total solar irradiance (TSI) 1978-2003 (ftp://ftp.ngdc.noaa.gov/STP/SOLAR_DATA/ SOLARJRRADIANCE/)

The results are shown in Figure 11. It is found that the autocorrelation method is inapplicable to the Sun, because the Carrington period of rotation is P=27.2753 days instead of $1 < P < 4$ days in our stellar set. In this case, the time scale of non-rotational variability of $F$ (due to spot decay processes, faculae variation, sub-photospheric convection, etc.) becomes shorter than $P$. Correspondingly, the solar one-period light-curves are highly noisy. Sometimes the rotational modulation is undetectable (Figures 11a). The rotational peak in periodograms does not dominate on the background of numerous spectral peaks (Figure 11b). Such variability of one-period light-curves decreases and distorts the autocorrelation $r_1(\Delta t/P)$ (Figure 11 d,j) leading to a positive coefficient $a$ in Equation (11), and therefore, the imaginary angular frequency $\omega_{cyc} = (-2a)^{1/2}$ of the stellar cycle.

### 3.4. Histogram method (HM)

As can be seen in Fig. 7a,c,e and 10c,i, the cycle period is the time interval between adjacent minima of the activity index, e.g., $A_1^2(t)$. Unfortunately, usually the fixing of such minima is not an easy task because of noisy irregularities of the curves $A_1^2(t)$, (see for example in Fig. 10 i). To avoid this difficulty, we construct the histogram of time intervals between stellar rotations, where the condition $A_1^2(t) < \langle A_1^2 \rangle / 4$ is fulfilled. The examples of such histograms are shown in Figure 9e,m. One can see the oscillations of such histograms, which reproduce the stellar cycle periods. Finally, we measure the cycle period $P_{HM}$ as a time interval between the major maximum (at $\Delta JD = 0$) in the histogram and first adjacent maximum, marked with arrows in the middle-column plots in Figure 9e,m. To decrease the probability of erroneous estimates, we did not consider the histograms with low oscillation contrast, i.e., with $n_{min}/n_{max} > 0.7$, where $n_{min}$ is the number of which corresponds the first minimum (closest to $\Delta JD = 0$) and $n_{max}$ is the number of $P_{HM}$-estimates at the next maximum. The width of the histogram bin is accepted as a probable error $\sigma_{HM} \sim 0.1 P_{HM}$.

This method is tested and tuned using simulated light-curves (see Sect. 3.3). Obtained values of the cycle period $P_{HM}$ are correlated with $P_{ACM}$ and the peaks in power spectra of $A_1^2$ (Figure 9f,n). The distribution of log($P_{ijM}$) is gaussian-like with maxima at the model value (Figure 9h,p). Its mean value $\langle \log(P_{ACM}) \rangle$ corresponds to $P_{HM} = 51.4 \pm 0.3$ and $285.1 \pm 5.7$ days, which are close to the model input parameter $P_{cyc} = 50$ and 300 days.

Figure 12 shows that the peaks (arrowed) in HM histogram detects the well-known periodicities in solar TSI data: the 11-yr cycles and the Rieger period of 150 ± 17 days (Ballester et al, 1999; Richardson & Cane, 2005 and therein). Note that the used activity index $A\backslash$ reveals the short-period Rieger-component of solar activity as oscillations with the cycle of 100-200 days in Figure 11e. Figure 11f demonstrates that HM is sensible to the short-period variations of solar activity in spite of dominating 11-yr cycle.

Figure 12 shows the statistical accordance between results of the histogram and autocorrelation methods. Since the methods confirm each other, they are used for the study of stellar activity.

## 4. Analysis of the results

Each of the above presented methods for $P_{cyc}$ estimation has advantages and disadvantages. For example, ACM deals with the second order effect (i.e., quadratic term in Equation (11)), which is mainly responsible for the larger scattering of the period estimates (compare histograms (g) and (h) or (o) and (p) in Figure 9). On the other hand, the ACM approach has broader applicability as compared with the HM approach. That is why below we use both approaches. Individual errors of each method could be decreased statistically by averaging over the large number of estimates.

Inevitably, there is statistical noise in estimates of the coefficients in Equation (11) giving sometimes positive $a$ and therefore imaginary $\omega_{cyc} = (-2a)^{1/2}$ of the stellar cycle in some models and real cases (e.g., KIC 6056983 in Figure 7). Moreover, some of the stars demonstrate rather irregular variations of their activity level (e.g., Figure 7e) giving too low contrast of the peaks in histograms like in Figure 9k for HM. That is why the application of the above described methods to 513 light-curves enables us to measure $P_{cyc}$ in 90% of cases (462 objects).

We analyze the measured values of $P_{ACM}$ and $P_{HM}$, keeping in mind two possibilities regarding their nature. First, these could be indeed the manifestations of the short stellar activity cycle related to the operation of the stellar magnetic dynamo (Vida et al. 2014). Alternatively, some periodicities of the activity indexes may be caused by the effects of beating due to differential rotation of the surface activity complexes at different latitudes

(Mathur et al. 2014). Below we consider both these options to decide on the best interpretation of the obtained results.

### 4.1. Phenomenology of short-period cycles

Mathur et al. 2014, analyzing two stars (KIC 3733735 and KIC 9226926), interpreted the measured values of the light-curve short periodicity $(P_{cyc} \sim 100$ days) as a result of beating of two rotational frequencies of two spots (or active regions) at different latitudes. While the proposed interpretation seems to work in the particular cases, it is not universal. Below we provide arguments in support of this point.

1. We search for hidden signs of probable beatings in stellar light-curves. To reveal such diagnostic features of beatings, we consider the pair of starspots with different periods $p_1 = P - \Delta P/2$ and $p_2 = P + \Delta P/2$ of stellar rotation at latitudes $\varphi_1$ and $\varphi_2$. The corresponding light-curve with a beating can be modeled using the following equation

$$F(t_i) = 1 - \sum_{j=1}^{2} G_j \Psi_j, \qquad (17)$$

where $G_j$ is the area of $j$-th spot relative the total stellar disk, and $\Psi_j$ is the geometrical factor with $\Psi_j = \cos(\beta_j)$ or $\Psi_j = 0$ at the spot visibility or invisibility respectively. Here, $\beta_j$ is the angle between the surface normal at the spot and the observer direction. One can find, using the cosine rule of spherical trigonometry, that

$$\cos(\beta_j) = \cos(I)\sin(\phi_j) + \sin(I)\cos(\phi_j)\cos\left(\frac{2\pi}{p_j}t_i\right), \qquad (18)$$

where $I$ is the inclination angle between the rotation axis of the star and the observer direction. The spot is invisible (i.e., $\Psi_j = 0$) when $\cos(\beta_j) < 0$. Figure 13a is a fragment of the typical light-curve, which is calculated for the non-disappearing spots ($\varphi_1 = 40°$, $\varphi_2 = 50°$, $I = 30°$, $P = 3$ days, $\Delta P = 0.02P$ and $G_1 = G_2 = 0.01$). The processing of this light-curve, using the one-period method (Section 3.2), reveals that the amplitude $A_2$ of the second rotational harmonic oscillates in antiphase relatively to $A_1$ of the first harmonic (Figure 13b). Correspondingly, the strong anti-correlation coefficient $r_{12} \approx -1$ has been found between $A_1$ and $A_2$ independently from the model parameters and inequality between $G_1$ and $G_2$, but only in the case of non-disappearing spots. This effect is a result of the alternation of spot conjunction and opposition in longitude. The temporal invisibility of the spots (I= 60° in Figure 13c) forms the saturated maxima of the light-curve, and therefore, another pattern of $A_2$ in Figure 13d. The application of Fourier transform to such light-curves reveals another diagnostic property of beatings: $P_1/P_2 =2$ which is valid for the case of temporally invisibility of both spots at $—I< \varphi_j <I$. To reveal these beating signs (i.e., $r_{12} \approx —1$ and $P_1/P_2=2$) we have processed all light-curves of our data set and constructed the real distributions of $r_{12}$ and $P_1/P_2$ (Figure 14). The histogram of $r_{12}$ (Figure 14a) has a gaussian-like shape with a slight shift towards the positive correlation. No significant secondary maximum is visible at $r_{12}<0$. The ACM- and HM-histograms do not show any secondary peaks at $P_1/P_2=2$. Therefore, the beating effect is not a dominating one above activity cycles in the analyzed light-curves.

2. Figure 15 shows that our measured values of $P_{ACM}$ and $P_{HM}$ agree well with the measurements of $P_{cyc}$ for fast rotating *(P~1 day)* stars published in Vida et al. (2014). It is important to note that Vida et al. (2014) in their study used the time-frequency method, which is insensitive to the beating effect. Although the measurements in Vida et al. (2014) concern only the cold stars with $T_{eff}<4303$ K, the clustering of all our estimated periods for the broader temperature range of stars (including also the cold stars) in Fig. 15 has a form of a clear sequence. That gives us a reason to expect that our approach reveals a general relation between the stellar temperature and activity cycle length.

4. The obtained estimates of the normalized cyclic frequency $\omega_{cyc}/\Omega=P/P_{cyc}$ of the stellar cycles are clustered mainly in the unstudied region of the SB diagram (see Figure 16). This clustering is extended from the region of inactive (I) stars towards the branch of super-active (S) objects. This effect is especially well pronounced in Figure 16b which represents the stars with the shortest periods ($1<P<1.5$ days), hence, with the best autocorrelation statistics within the available *Kepler* observation time frame. A similar connection has been addressed also for the super-active (S) and active (A) star branches (Saar & Brandenburg, 1999). Therefore, our measured cycle periods and corresponding $\omega_{cyc}$ values show the analogous behavior in SB diagram as the known and recognized cycles do. This again gives us a reason to argue that the obtained estimates of $P_{ACM}$ and $P_{HM}$ characterize the real activity cycles.

## 4.2. Interpretation of the results in terms of stellar dynamos

The classical αΩ-dynamo theory (Parker, 1955) predicts the angular velocity $\omega_{cyc}$ = $2\pi/P_{cyc}$ of the activity cycle as

$$\omega_{cyc} \approx \sqrt{\frac{1}{2}|\alpha S k|}, \qquad (19)$$

where $\alpha$ is a parameter which describes the transformation of the toroidal magnetic field into the poloidal one, $S$ is the radial gradient of the angular velocity of stellar rotation and $k$ is the wavenumber of the dynamo-wave.

To avoid the exponential growth of the magnetic field in the dynamo models, a sharp zeroing of the parameter $\alpha$ (cut-off in $\alpha$ or α-quenching) was proposed above a critical threshold of the magnetic field strength (Robinson & Durney, 1982, and therein). More realistic variations of the parameter $\alpha$, depending on the energy density of the magnetic field, until now remains a matter of debate (Brandenburg & Subramanian, 2005, and references therein). This discussion involves heuristic arguments, theoretical modeling and numerical experiments. At the same time, Equation (19) can be used in the experimental approach to the problem of α-quenching.

The generation of magnetic fields is commonly parameterized by the dynamo number $N_D$, which is essentially the ratio between the magnetic field generation and diffusion terms in the convective zone. Generally, the theory of αΩ-dynamo predicts that $N_D \approx \text{Ro}^{-2}$ (e.g., Noyes et al., 1984). Hence, the increase of $\text{Ro}^{-1}$ in Fig. 16 leads to the increase of $N_D$, which means the growth of the magnetic field in the convection zones of stars (see, e.g., Fig. 1 in Robinson & Durney, 1982).

Sufficiently strong magnetic field can affect the flow of plasma and suppress the dynamo process. This effect is known as α-quenching which decreases the parameter $\alpha$, and hence the cycle frequency $\omega_{cyc}$ according to Equation (19). Our measurements shown in Figure 16, clearly indicate the decrease of $\omega_{cyc}$ for the increasing Ro$^{-1}$. One can also see in Fig. 7, 10 and 15 the increase of $P_{cyc}$ for the decreasing $T_{eff}$ of the stars (i.e., increasing $\tau_{MLT}$ and Ro$^{-1}$).

However, the stars with lower Ro$^{-1}$ show the opposite behavior in Figure 16. According to Noyes et al. (1984b), that contradicts the α-quenching effect. The difference in the slopes of the I-branch and the most of our measurements cluster in the SB diagram can be attributed to the α-quenching effect for fast rotating stars (1<P<4 days).
An additional argument for the α-quenching in our stellar set follows from Fig. 17. Our estimates show the abrupt decrease of $\omega_{cyc} \propto \alpha^{1/2}$ (see Equation (19)) with the increasing $A_1^2$, i.e. with the growth of stellar activity, hence, the magnetic energy of a star. This effect is visible in the theoretical plot (Fig. 17a; based on the results by Rüdiger & Kichatinov, 1993). On the other hand, the beating period is a result of spot difference in longitude, independently from the spot sizes (i.e., related $A_1^2$). Therefore, the clear quenching-like dependence between $\omega_{cyc}$ and $A_1^2$ is an additional argument against the beatings as a main cause of short cycles.

The effect of α-quenching controls a saturation level of the stellar activity. This fact enables another verification of our interpretation of the obtained results in terms of α-quenching. Wright et al. (2011) find that the coronal X-ray saturation in solar- and late-type stars occurs at Ro < 0.13 ± 0.02. Taking into account the difference between definitions of Rossby number Ro=$P/\tau_{MLT}$ in Wright et al. (2011) and $Ro= P/(4\pi \tau_{MLT})$ in Saar & Brandenburg (1999), the saturation region, found by Wright et al. (2011), transforms in the SB diagram to the area of Ro$^{-1}$ >2. One can see in Figure 16 that the cluster of our estimated values of $\omega_{cyc}/\Omega = P/P_{cyc}$ declines just at Ro$^{-1}$≈2 from the I-branch in the SB diagram and demonstrates the decreasing trend. This once again confirms the manifestation of the α-quenching effect in our measurements.

# 5. Conclusions

To our knowledge the study presented above is the first attempt to automate the process of measuring of the characteristic periods of stellar cycles. This task is not a trivial one because of the complexity of stellar activity variations with different interferences especially in the cases of non-dominating short-period components.

However, namely such short-period cycles are particularly interesting, because they are supposed to be a manifestation of the dynamo process at the bottoms of convection zones (Bohm-Vitense, 2007), which dominates in the Sun.

That is why our methodology was especially dedicated to deal with the short-period component of stellar activity. The application of these methods to the *Kepler* photometry enables us to draw the following conclusions.

1. The automated measuring of short-period component of rotational modulation of stellar light-curves gives information on real activity cycles. This opens a way for the regular survey of activity cycle periods.

2. The studied fast-rotating stars have the cycles which form the previously unknown sequence on the SB diagram, connecting the region of inactive stars with the branch of super-active objects.

3. The existence of this new sequence is a result of the α-quenching effect, which decreases the cycle periods in the dynamo saturation regime at $Ro^{-1} > 2$. It is important in the context of the discussion on catastrophic quenching and for other heuristic approximations of the α-quenching effect (Hubbard & Brandenburg, 2012).

The proposed approach and methods will be used for a more extensive survey of stellar activity of slow as well as super-fast rotators among the stars of the Main Sequence and giants.

This work was performed as a part of the project P25587-N27 of the *Fonds zur Forderung der wissenschaftlichen Forschung, FWF*. The authors also acknowledge the *FWF* projects S11606-N16, S11604-N16, S11607-N16. This research has made use of the NASA Exoplanet Archive, which is operated by the California Institute of Technology, under contract with the National Aeronautics and Space Administration under the Exoplanet Exploration Program.

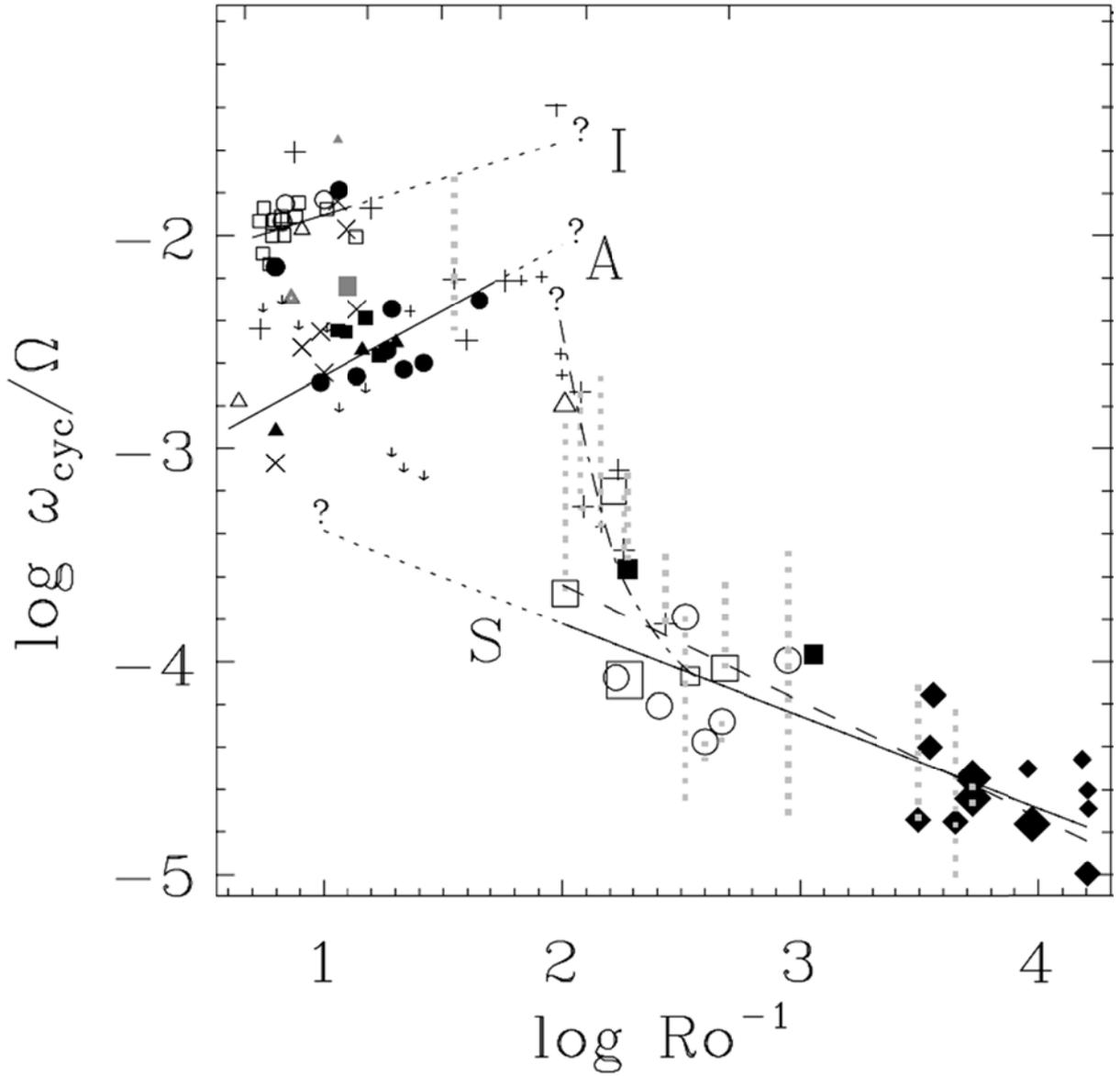

Fig. 1.— The SB diagram (Fig. 5 in Saar & Brandenburg, 1999) shows the relations between the normalized angular velocity of the stellar cycles $(\omega_{cyc}/\Omega)$ and the inverted Rossby number $(Ro^{-1})$ for the inactive (I), active (A) and super-active (S) stars. The questioned continuation of the I-branch is the subject of our study.

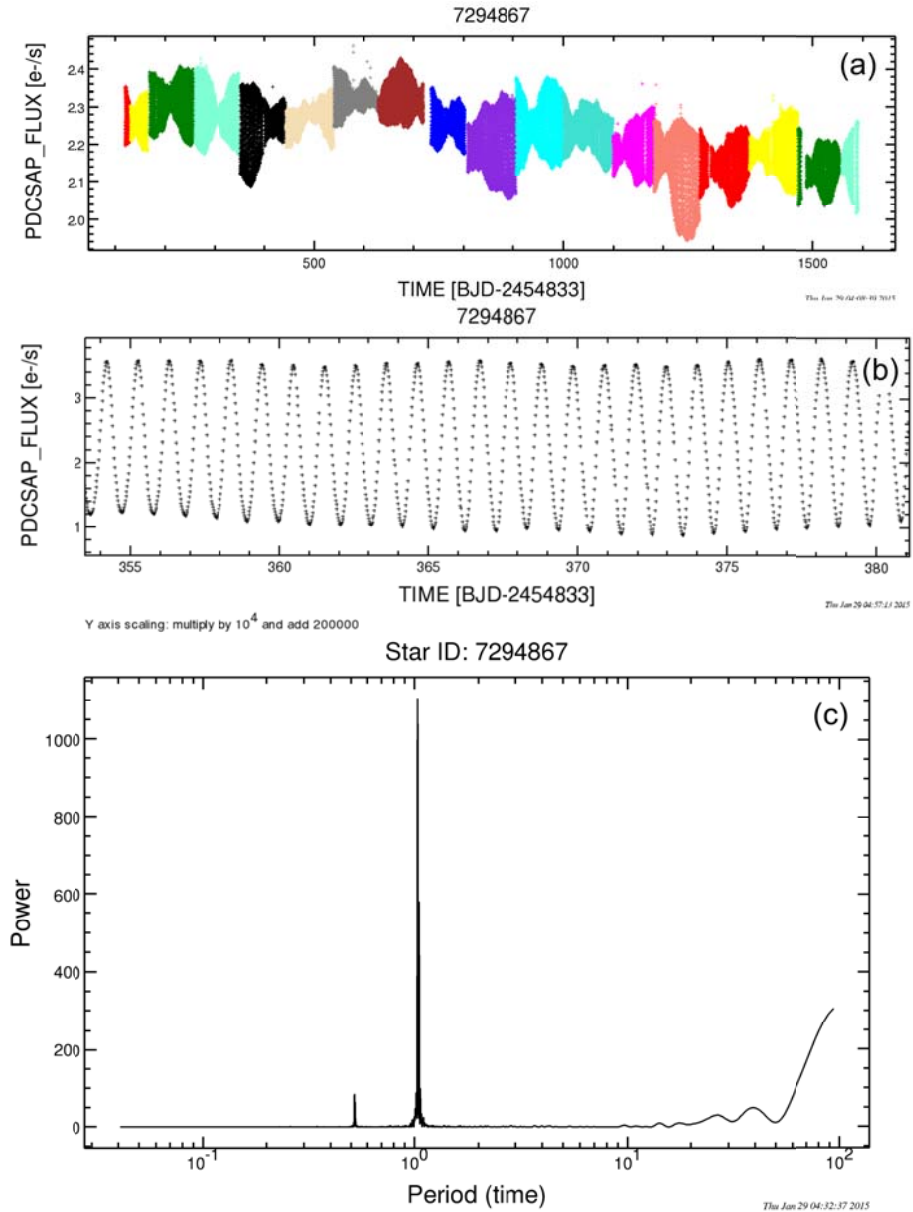

Fig. 2.— The light-curve of KIC7294867 demonstrates attributes of the spot-related rotational variability which has been used for selection of the sample of stars: a) wave-like variations of the variability amplitude (data for all observation quarters are shown using different colors); b) rotational oscillations with period 1< $P$ < 4 days (the fragment of quarter 5 in (a) is shown in details); c) domination of rotational harmonics in the Lomb-Scargle periodogram *(P, P/2)* at timescales shorter than month (data for quarter 5 were used). All plots were prepared using NASA Exoplanet Archives service (http://exoplanetarchive.ipac.caltech.edu/).

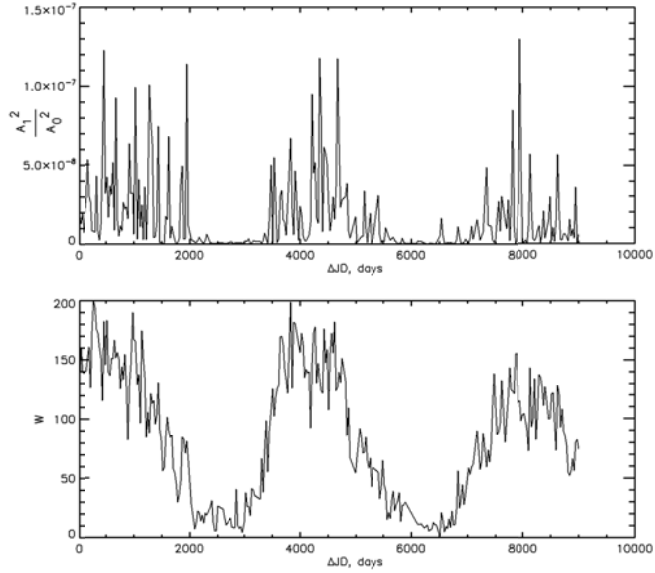

Fig. 3.— Solar activity cycles are shown using the activity index $A_1^2/A_0^2$ (TSI; top panel) and the classical total sunspot number $W$ (lower panel). The time scale is given in Julian days $(\Delta JD)$ counted from 16 November, 1978 (the starting day of measurements).

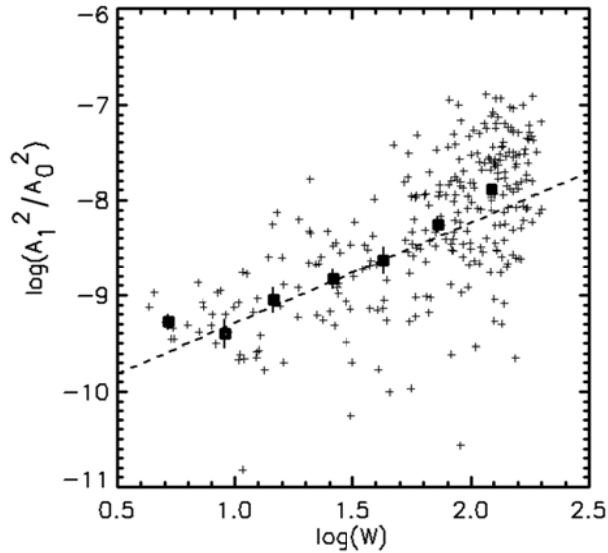

Fig. 4.— Correlation between solar activity index $A_1^2/A_0^2$ based on the TSI and the total sunspot number $W$ for Fig. 3. The activity level is measured in both indices for identical solar rotations (crosses). The averaged values (squares with error bars) reveal the quasi-linear relation with the regression (dashed line) $A_1^2/A_0^2 = GW + Q$, where $G = 1.17 \pm 0.08$ and $Q = -10.40 \pm 0.15$.

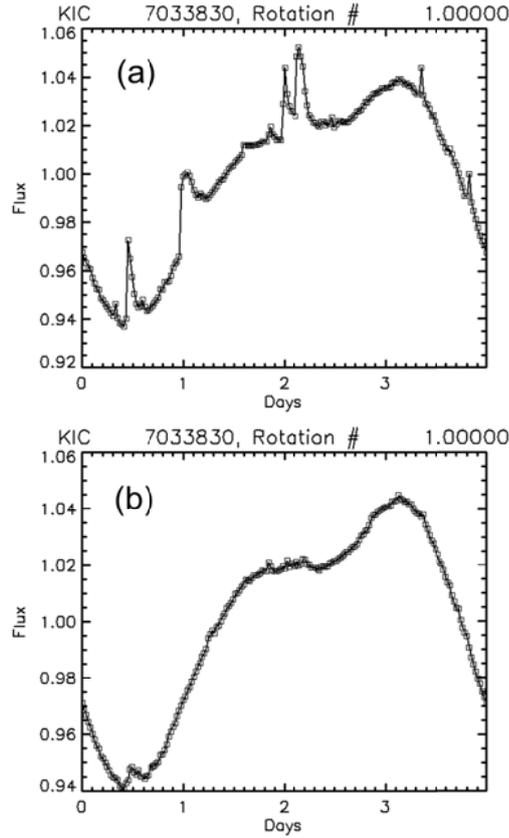

Fig. 5.— The one-period light-curve for the star KIC 7033830 (a) without and (b) with the correction of flares. For both curves, the trend is removed, and the flux is normalized $F/\langle F \rangle$, where $\langle F \rangle$ is the average flux for this time interval $P$.

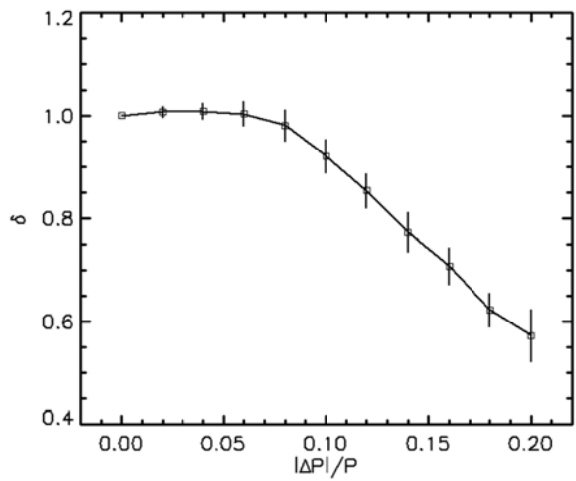

Fig. 6.— Influence of the error $\Delta P$ of the stellar rotation period $P$ on the squared amplitude $A_1^2$ of the first harmonic of rotational modulation of $F$ (the analyzed parameters in our method). The parameter $\delta = \langle A_1^2(P + \Delta P)/A_1^2(P)\rangle$ is the average ratio between the values $A_1^2$ calculated for the same stellar rotation but using the adopted $P$ and slightly modified $P + \Delta P$ rotation periods. The individual ratios were averaged (squares and solid line) using the whole stellar set (513 objects).

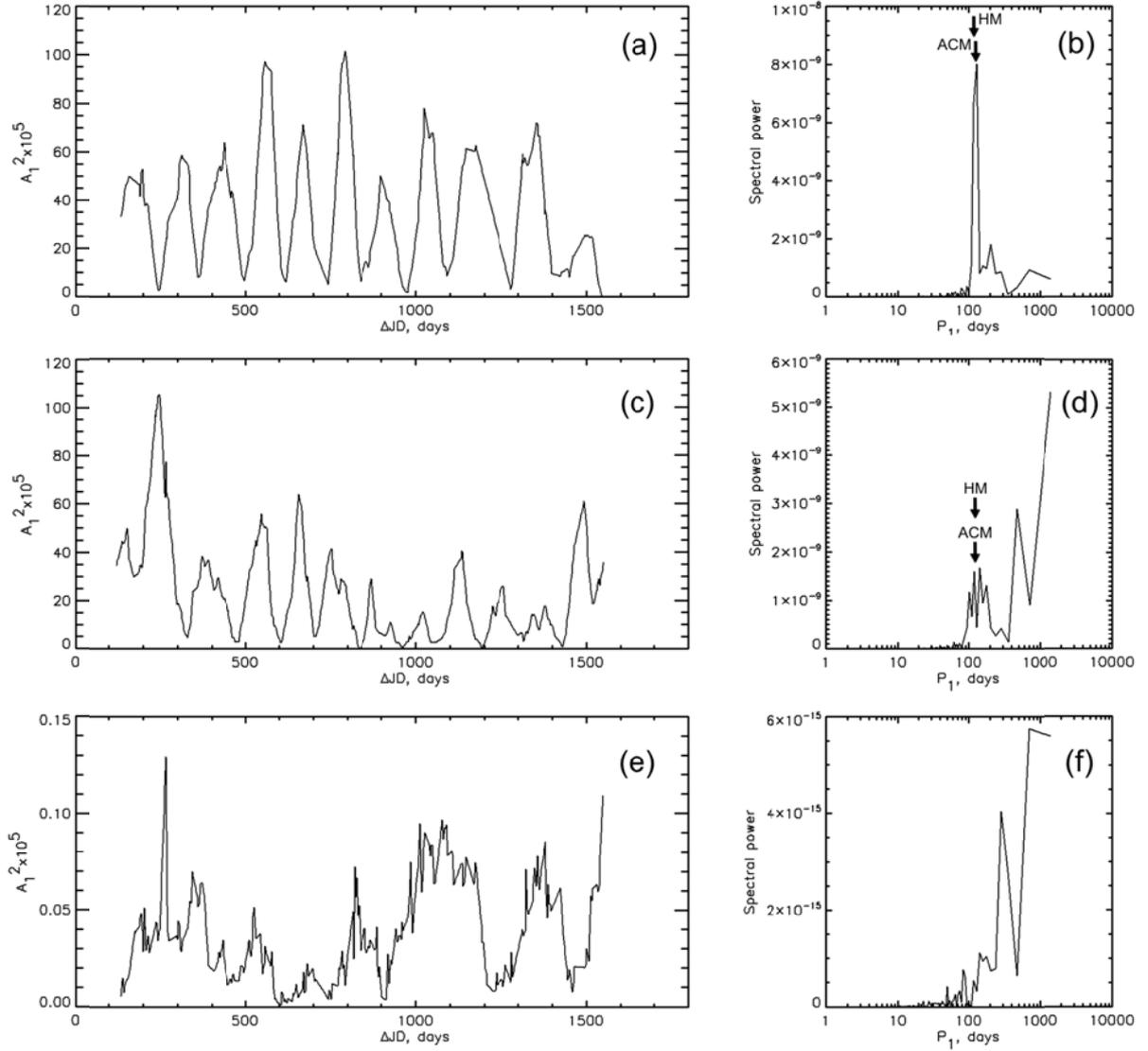

Fig. 7.— Examples of stellar activity variations, i.e., index $A_1^2$ (left-column plots) and its corresponding power spectrum after Fourier transform (right-column plots) for three sample stars: (a) and (b) for KIC 1570924 ($T_{eff}$ = 4923 K; $P$ = 3.244 days); (c) and (d) for KIC 4543412 ($T_{eff}$ = 5907 K; $P$ = 2.163 days); (e) and (f) for KIC 6056983 ($T_{eff}$ = 5719 K; $P$ = 2.902 days). The arrows in (b) and (d) show the cycle period, measured with the methods described in Sections 3.3 and 3.4.

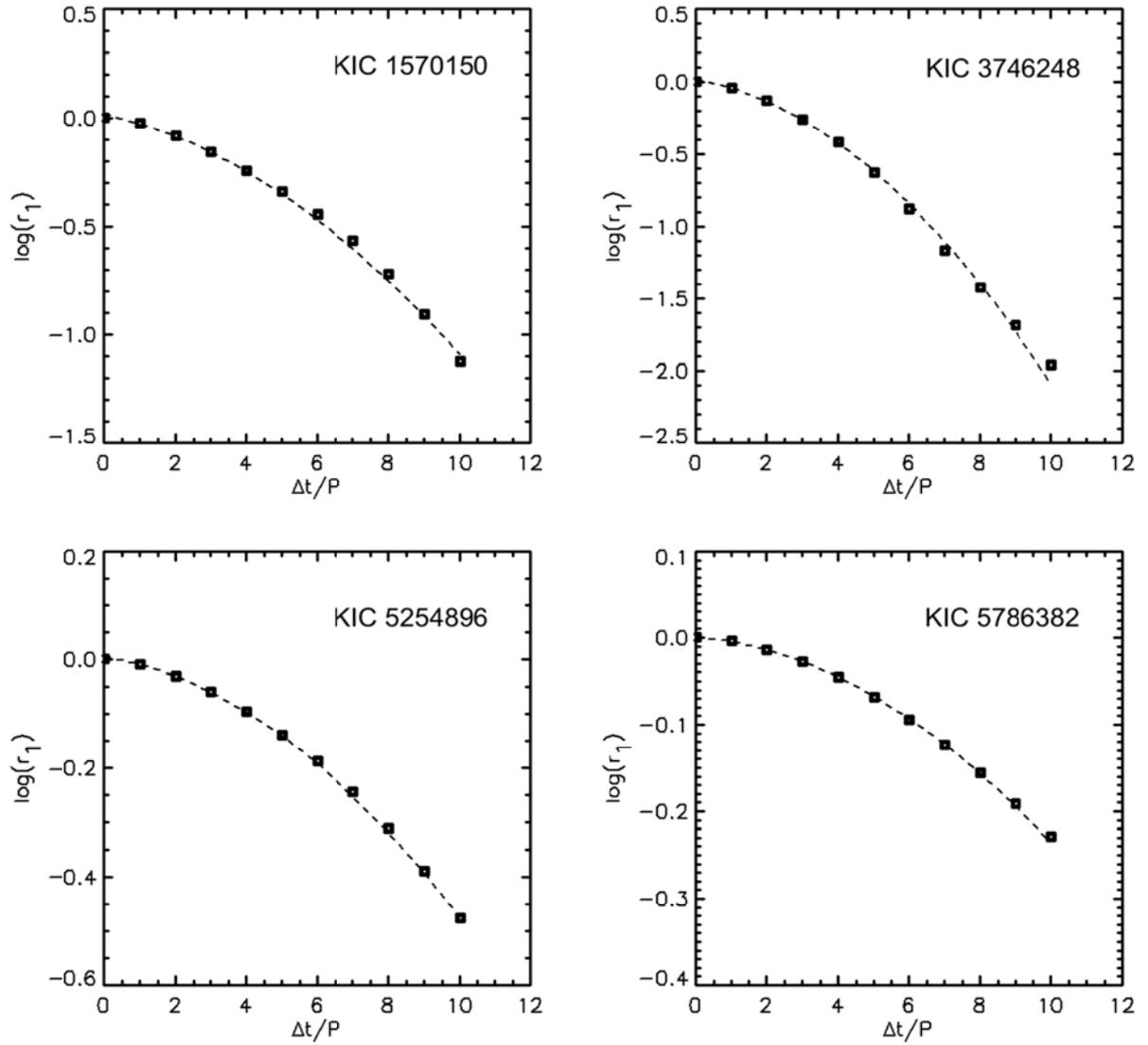

Fig. 8.— Examples of quasi-quadratic autocorrelation of $A_1^2$. The experimental values of autocorrelation coefficient $r_1$ (squares) are approximated with quadratic polynomials using the lags $1 < \Delta t/P < 4$ for every star (labeled).

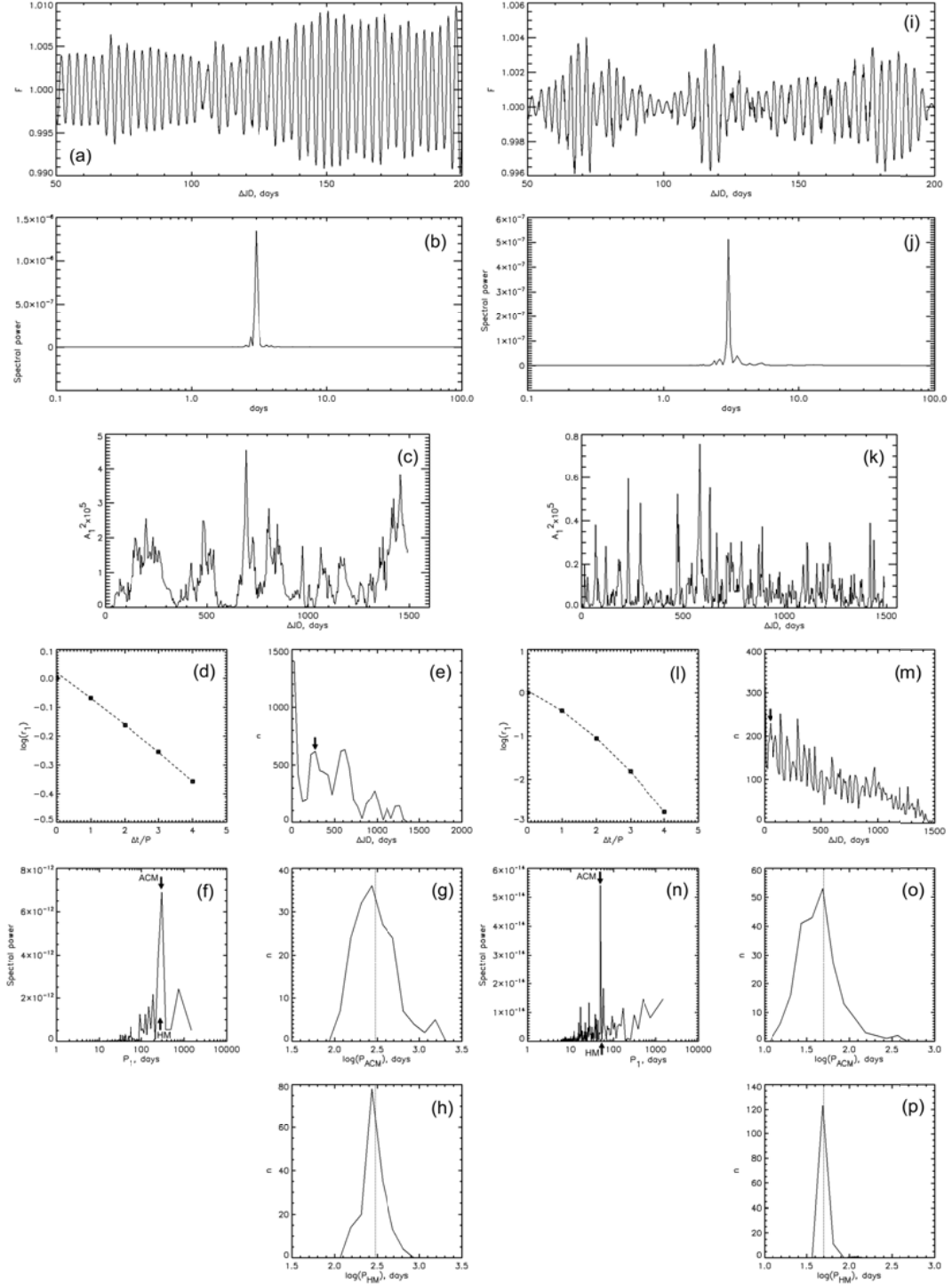

Fig. 9.— Examples of processing of the simulated light-curves for the cold $(T_{eff} \sim 3500$ K) and hot $(T_{eff} \approx 6700$ K) stars are shown in panels (a)-(h) and (i)-(p), respectively: (a) and (i) are the fragments of synthetic light-curves; (b) and (j) show the corresponding periodograms; (c) and (k) are the variations of the activity index $A_1^2$ with the time $\Delta JD$ (days counted from the simulation start); (d) and (l) depict the autocorrelation curves of $A_1^2$ in (c) and (k) with the quadratic approximation curve; (e) and (m) show the histograms of time intervals between the measurements, when $A_1^2(t) < \langle A_1^2 \rangle / 4$ (see Sect. 3.4); (f) and (n) are the comparisons between the found $P_{cyc}$ (arrows indicate particular method) and the power spectra (Fourier transform) of activity curves in (c) and (k); (g) and (o) are the distributions of cycle period measurements for ACM in comparison with the specified $P_{cyc}$ in Equation (16) (the dotted verticals); (h) and (p) are the histograms analogous to (g) and (o), but obtained with HM; $n$ is the number of the period measurements in a bin of the histogram.

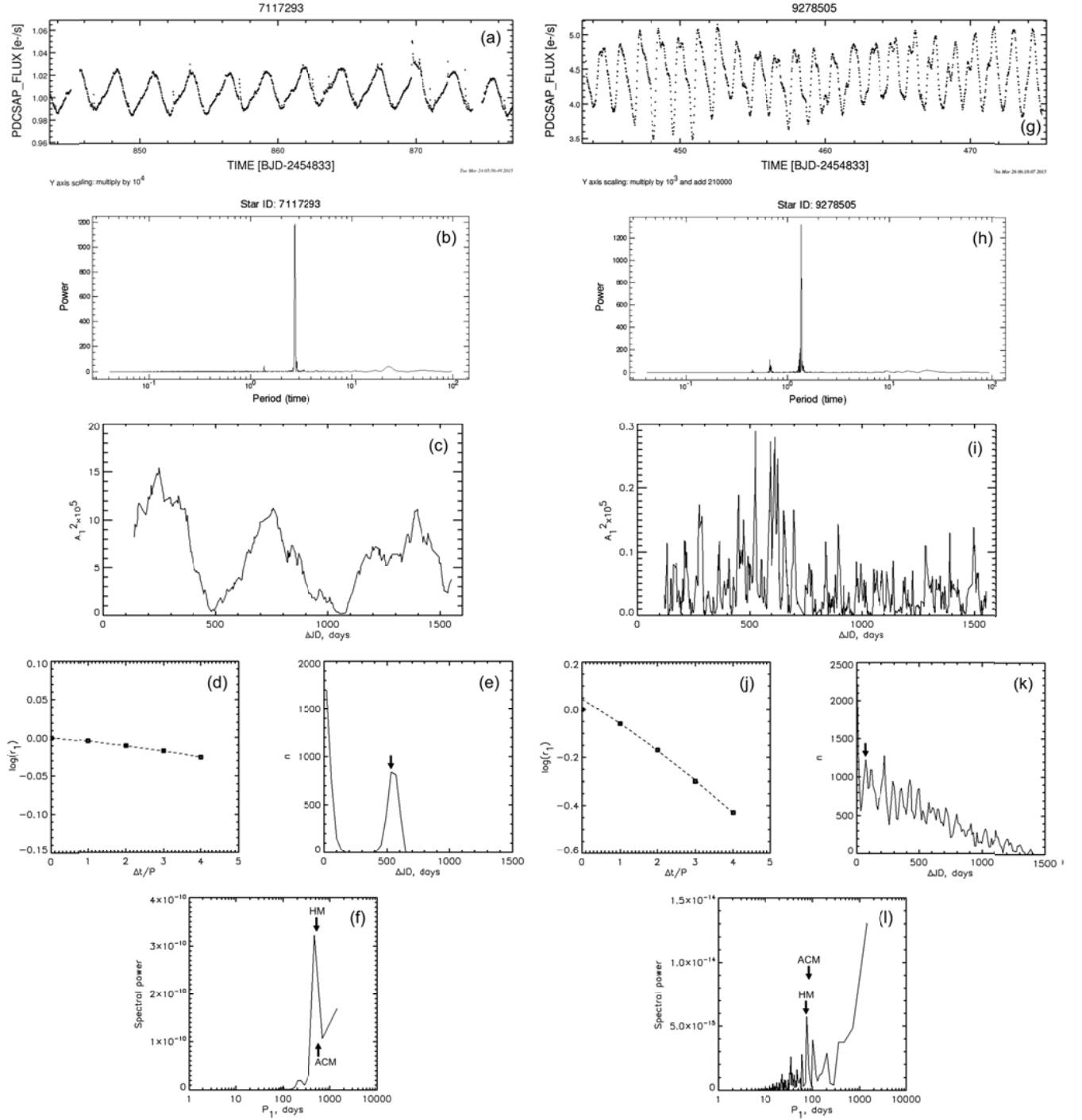

Fig. 10.— Applications of the proposed methods (ACM and HM) to the real light-curves of KIC 7 117293 ($T_{eff}$ = 3621 K) - panels (a)-(f), and KIC 9278505 ($T_{eff}$ = 6511 K) - panels (g)-(l): (a) and (g) are the fragments of light-curves; (b) and (h) show the corresponding periodograms; (c) and (i) are the variations of the activity index $A_1^2$ with the time $\Delta JD$; (d) and (j) depict the autocorrelation curves of $A_1^2$ in (c) and (i) with the quadratic approximation curve; (e) and (k) show the histograms of time intervals between the measurements, when $A_1^2(t) < \langle A_1^2 \rangle/4$ (see Sect. 3.4); (f) and (l) are the comparisons between the found $P_{cyc}$ (arrows indicate particular method) and the power spectra (Fourier transform) of activity curves in (c) and (i). The plots (a), (b), (g) and (h) were prepared using NASA Exoplanet Archives service (http://exoplanetarchive.ipac.caltech.edu/).

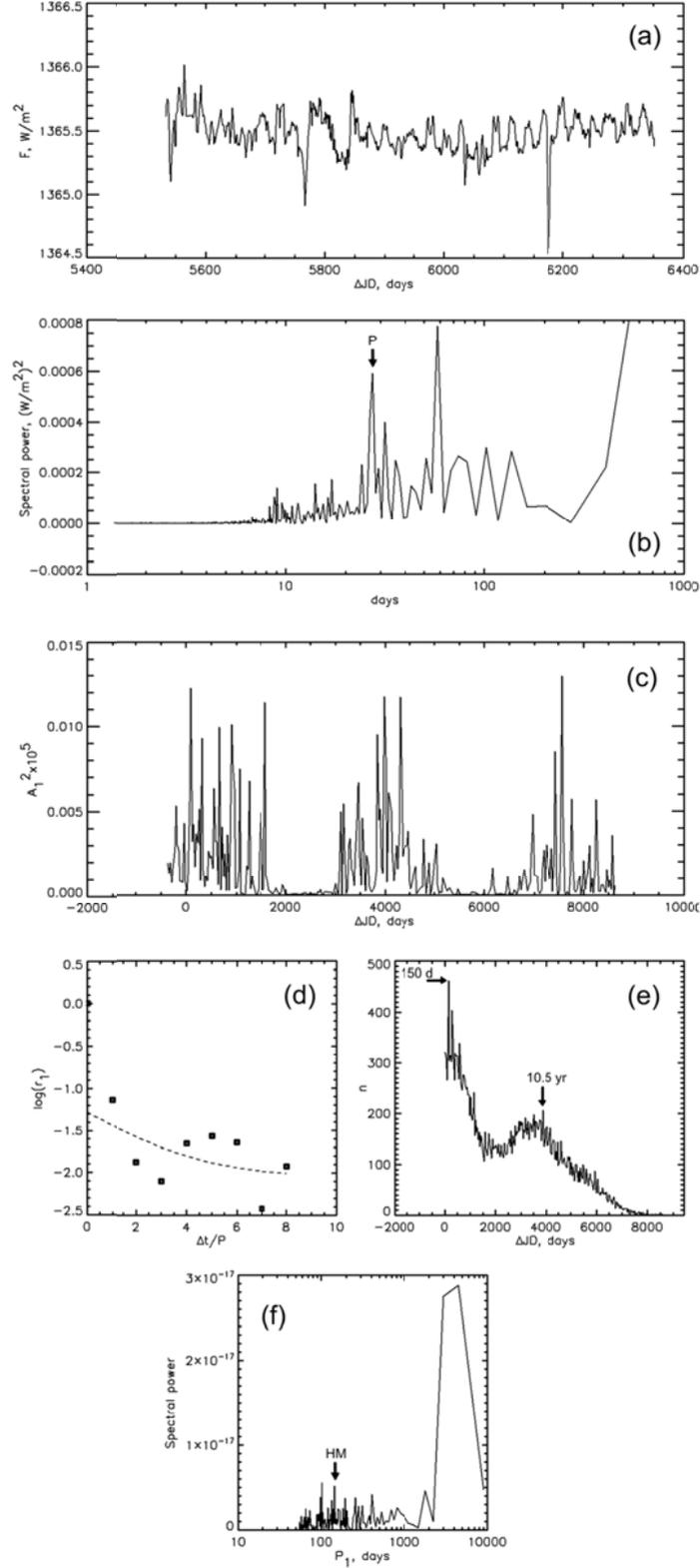

Fig. 11.— Applications of the proposed methods (ACM and HM) to the solar light-curves using TSI 1978-2003: (a) is a fragment of the light-curves; (b) shows the corresponding periodogram (the rotation period $P$ is arrowed); (c) is the variation of the activity index $A_1^2$ with the time $\Delta JD$; (d) depicts the autocorrelation curve of $A_1^2$ in (c) with the quadratic approximation curve; (e) shows the histogram of time intervals between the measurements, when $A_1^2(t) < \langle A_1^2 \rangle/4$ (see Sect. 3.4); (f) is the comparisons between the found $P_{cyc}$ (arrow indicates the method) and the power spectrum (Fourier transform) of activity curve in (c).

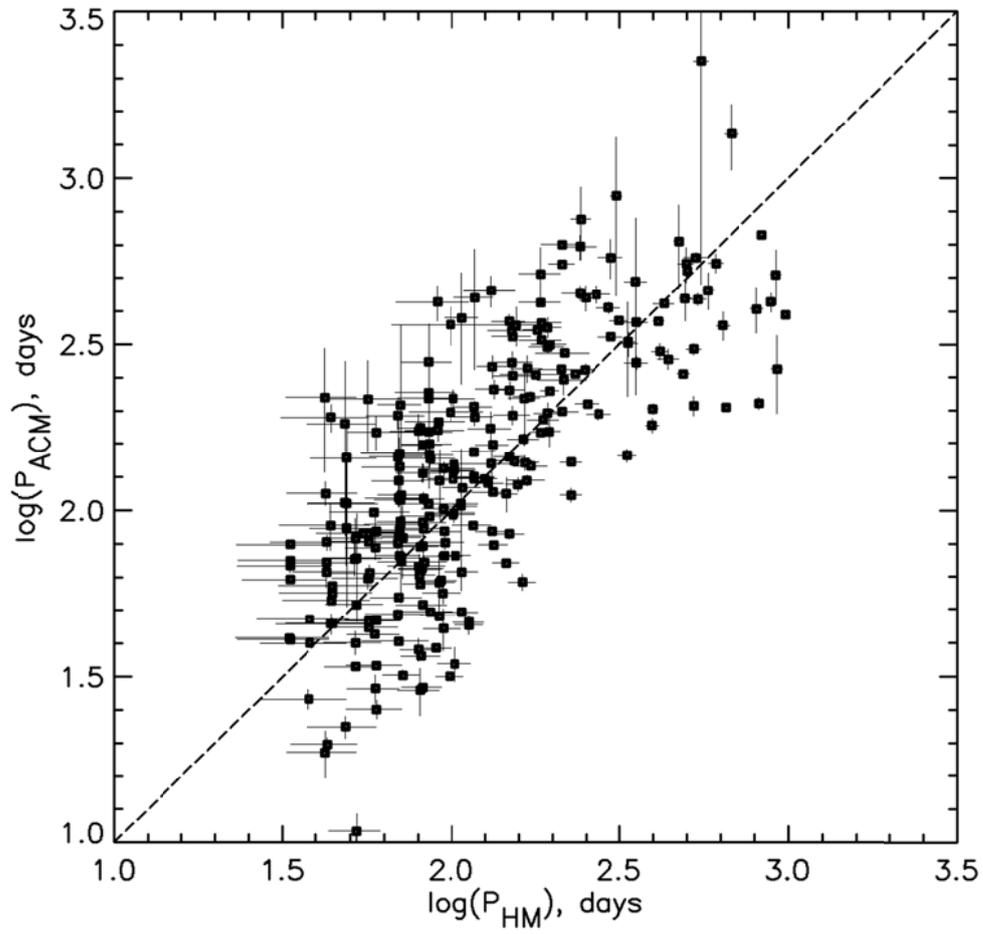

Fig. 12.— The stellar cycle periods, measured using both methods (indexes AMC and HM, respectively). The dashed line corresponds to the equality of estimates. The horizontal and vertical bars are the corresponding errors in $P_{ACM}$ and $P_{HM}$.

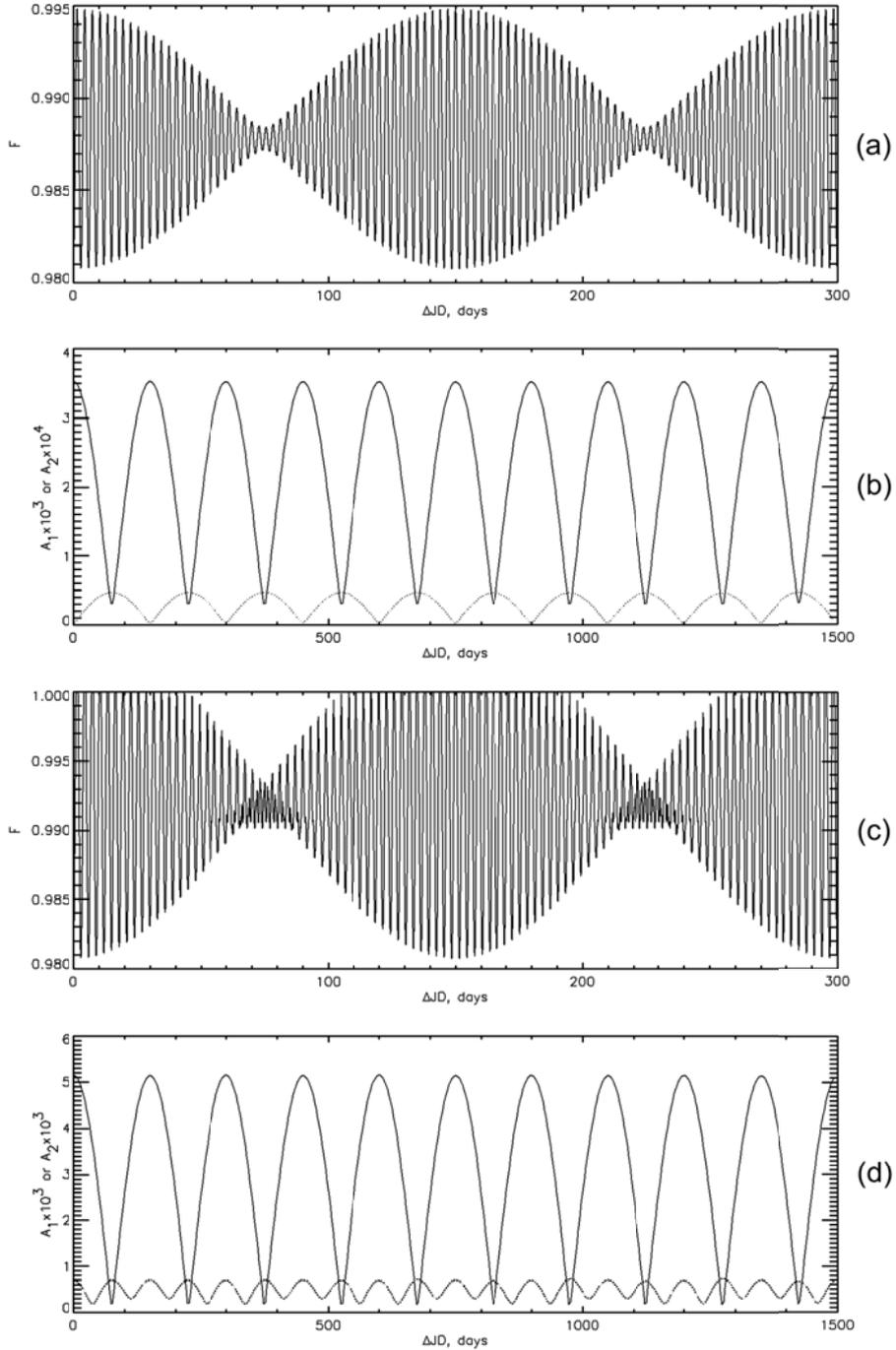

Fig. 13.— Beating phenomenology: (a) and (c) are sample fragments of the light-curves, which are synthesized using Equation (17); (b) and (d) are the corresponding variations of the amplitudes of the first $A_1$ (solid curve) and second $A_2$ (pointed) harmonics of one-period *(P)* light-curves. The plots (a) and (b) represent the case of non-disappearing spots *($\varphi_1=40°$, $\varphi_2=50°$, $I=30°$)*, while the graphs (c) and (d) show the effects from the temporal invisibility of the spots (*$I=60°$*).

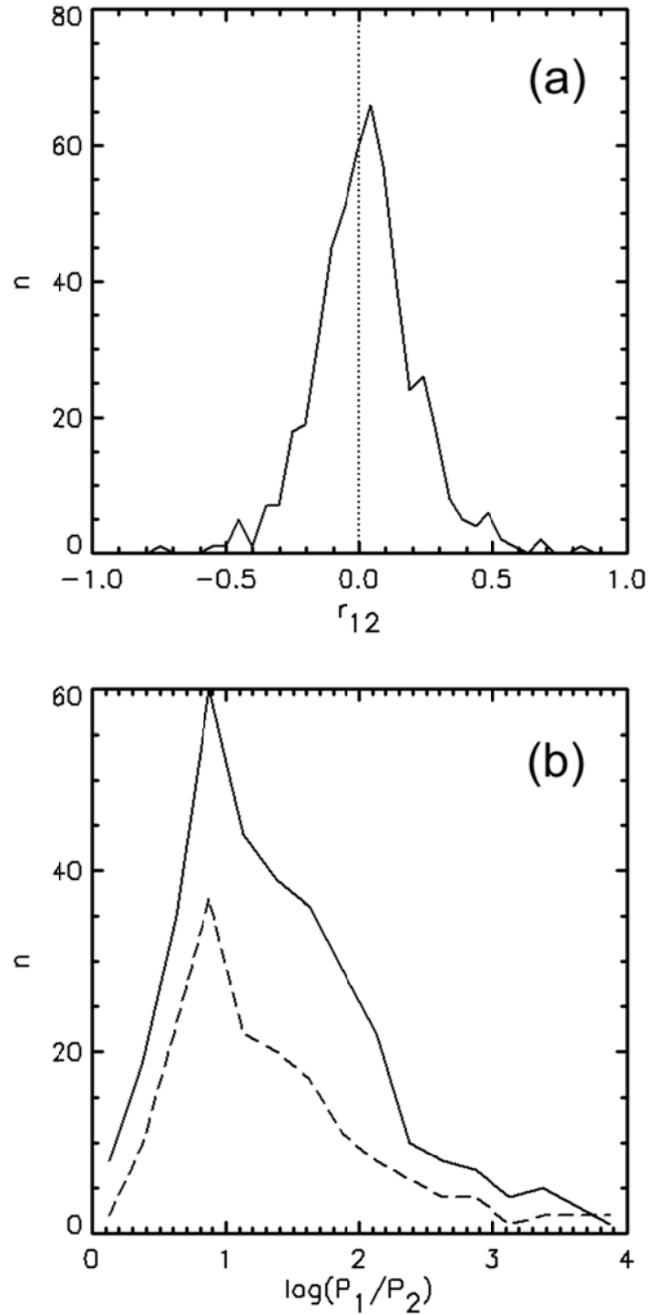

Fig. 14.— Search for the beatings in stellar light-curves: (a) the distribution of the correlation coefficient $r_{12}$ between the amplitudes $A_1$ and $A_2$ of the first and second rotational harmonics; (b) the histograms of the ratio $P_1/P_2$ between periods of $A_1$ and $A_2$ variations, which were found using ACM (solid line) and HM (dashed).

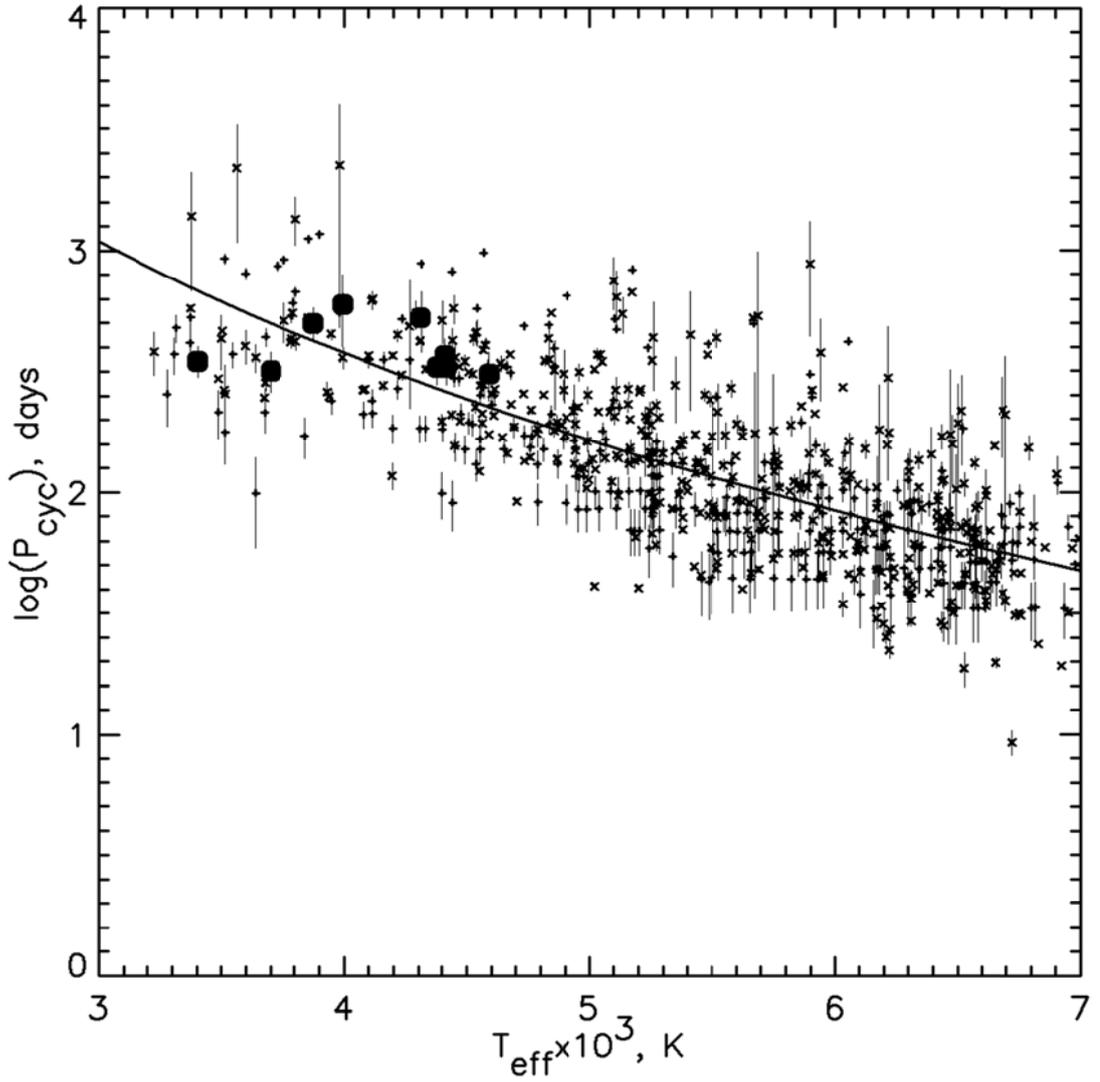

Fig. 15.— Cycle periods $P_{ACM}$ (marked with ×) and $P_{HM}$ (marked with +) versus the revised effective temperature of the stars $(T_{eff})$ according to Huber et al. (2014). Bolded points indicate the measurements by Vida et al. (2014) for the fast rotating red dwarfs $(P\sim1$ day). The approximation $\log(P_{cyc}/1\ \text{day}) = H \log(T_{eff}/1000°) + Z$ for all data is shown using the solid curve $(H=-3.70 \pm 0.12$ and $Z=4.80 \pm 0.09)$.

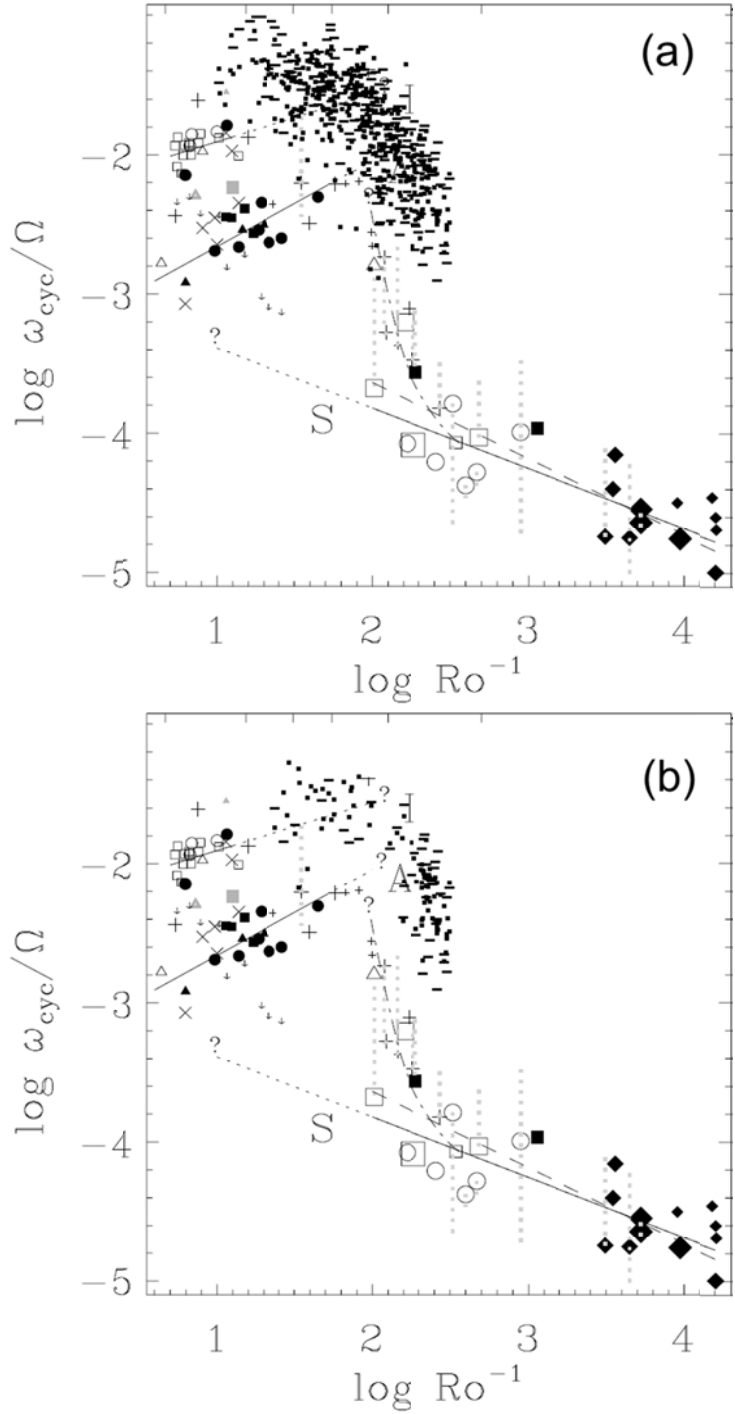

Fig. 16.— Estimates of $P_{ACM}$ (points) and $P_{HM}$ (minuses) superimposed onto the SB diagram (Saar & Brandenburg, 1999), the original of which is shown in Fig. 1: (a) all estimates over studied rotation periods (1<$P$<4 days); (b) the estimates only for the stars with the shortest rotation periods 1<$P$<1.5 days, i.e., the best rotation statistics for analysis. The inverted Rossby number $Ro^{-1} = 2\tau_{MLT}\Omega$ is found using the stellar rotation period $P$ (Nielsen et al., 2013) and the turnover time $\tau_{MLT}$ according to the approximation in Saar & Brandenburg (1999) after transformation from the color scale into $T_{eff}$ (Flower, 1996).

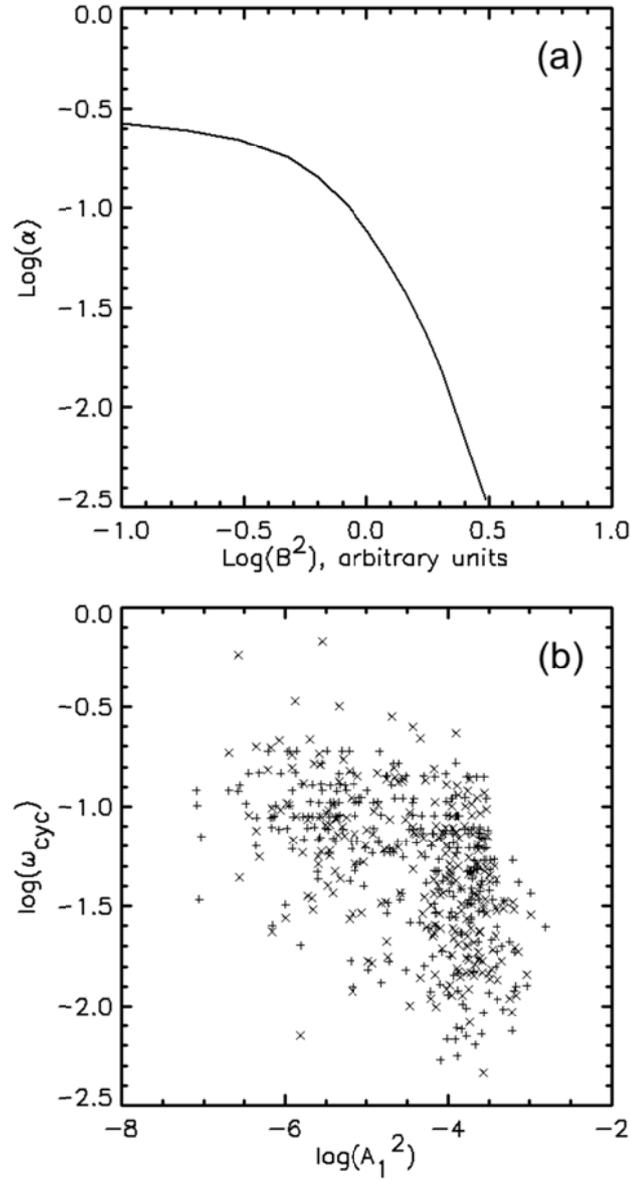

Fig. 17.— The α-quenching effect: a) the results of theoretical modeling (based on Fig. 2 in Rüdiger & Kichatinov, 1993, converted to the logarithmic scales) for the α-parameter as a function of the toroidal magnetic field energy; b) our experimental cycle periods $P_{ACM}$ (marked with ×) and $P_{HM}$ (marked with +) showing the abrupt decrease of $\omega_{cyc} \propto \alpha^{1/2}$ at $\log(A_1^2) \approx -3.5$. The increase of $A_1^2$ means the growth of stellar activity, i.e., the magnetic energy of a star.